# Mid-Air Haptic Bio-Holograms in Mixed Reality

Ted Romanus[1], Sam Frish[1], Mykola Maksymenko[1], William Frier[2], Loïc Corenthy[2], Orestis Georgiou[2]

[1] SoftServe Inc, Lviv, Ukraine
{first.last}@softserve.com

[2] Ultrahaptics Ltd., Bristol, BS2 0EL, UK
{first.last}@ultrahaptics.com

## ABSTRACT
We present a prototype demonstrator that integrates three technologies: mixed reality head mounted displays, wearable bio-sensors, and mid-air haptic projectors to deliver an interactive tactile experience with a bio-hologram. Users of this prototype are able to see, touch and feel a hologram of a heart that is beating at the same rhythm as their own. The demo uses an Ultrahaptics device, a Magic Leap One Mixed Reality headset, and an Apple Watch that measures the wearer's heart rate, all synchronized and networked together such that updates from the wristband dynamically change the haptic feedback and the animation speed of the beating heart thus creating a more personalised experience.

## CCS CONCEPTS
• **Human-cantered computing** ~ Haptic devices

## 1 INTRODUCTION

Mixed Reality (MR) headsets promise to revolutionize a plethora of applications ranging from entertainment to healthcare and have recently gained increased momentum and market traction. Already, these systems are extensively being used in enterprise type applications via serious games or simulators to train workforces. They may even be deployed in more challenging and complex environments such as warehouses [1], battle-fields [2] and hospital surgery rooms [3, 4]. In the latter case, MR brings to the healthcare, wellness and medical industry the opportunity for surgeons to visualize 3D models and better plan their operations. They can also communicate technical concepts to their peers and patients more effectively. For instance, prior to any operation, surgeons can obtain precise interactive 3D holograms of the body part undergoing surgery and with use of MR headsets, smoothly blur the demarcation between the physical world, medical scans and computer simulations. They may also interact in a 3D environment with the digital replicas of the organ that requires surgical treatment, view it from any desired angle or section, label it, analyse it, manipulate it using various hand gestures, share it with colleagues, and of course switch between 3D view and the traditional CT images. These interactions envisioned by current MR systems are however predominantly audio-visual and lack physicality or any other tactile information conveyed through our sense of touch.

Looking beyond the surgery room, and other localized immersive interactions, MR use cases are further empowered by advancements in wearable and sensor fusion technology and in wireless communications such as the tactile internet [5]. Traditional teleoperation, telemedicine, e-learning and remote interaction solutions such as voice or video conferencing, remote teaching, remote diagnosis and treatment, have reached a high level of sophistication and widespread use, thanks to the growth and progress of audio-visual communications and the commercial availability of affordable sensing hardware like smartphone cameras, GPS trackers, inertial measurement units (IMUs), photo-plethysmogram (PPG) and electrocardiogram (ECG) sensors, often found in fitness and wellbeing wearables and gadgets.

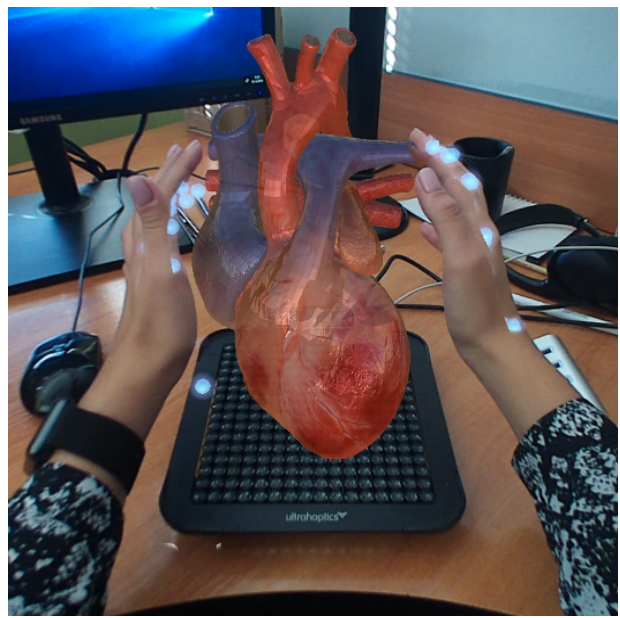

Figure 1: First person view of the MR headset during interaction with the haptic bio-holographic heart. Tracking markers are superimposed on the user's hands. Haptic feedback is activated when the markers intersect with the holographic heart.

With recent advances in communication technologies and the sensed data-integration fusion into MR systems, it is projected that a complete remote immersion can be realized with the added ability of physical tactile interaction with the remote environment. To that end, this can be achieved by the real time and wire-free exchange of multi-modal and bio-information, such as the combination of audio, video, haptic, environmental and bio-sensed information (e.g., heart rate), over the Internet or a local area network, and its rendering by a MR headset and an untethered haptic system. The inclusion of haptics in the above vision is paramount as it incorporates a completely new sensory dimension, thus improving the performance of the user interface [6] and enhancing the user's sense of agency [7] while also expanding the afforded interactions and deepening immersion levels. We should highlight that these benefits are application and implementation specific.

Within the Human-Computer-Interaction (HCI) field, haptic technology refers to devices able to create an experience of touch by interfacing with users through *force, vibration, or motion*. These devices target the human cutaneous and kinaesthetic sensory systems. This field borrows from many areas, including robotics, experimental psychology, biology, computer science, system and control, and others. Haptic devices have demonstrated significant advancements and are no longer limited to just hand-held controllers that vibrate within a limited set of degrees of freedom. Possible peripheral alternatives include especially designed controllers, haptic wearable gloves, electrical muscle stimulation pads, and even full body exoskeleton suits [8]. While traditional grounded haptic devices have had a significant impact in the last decade in the fields of robotic surgery and 3D design, they are less suitable for natural hand interactions than those facilitated by state-of-the art commercial MR headsets that employ inside-out multi finger hand tracking.

An alternative to the above touch interfaces is *touchless ultrasonic haptic technology* [9]. This technology employs electronically controlled phased arrays of ultrasound speakers (or transducers) to create high acoustic pressure points in mid-air that can be felt with bare hands. The technology has made significant leaps forward since its invention in 2010 [10]. It delivers advanced, multi-point mid-air tactile sensations directly onto users' hands and fingertips, allowing accurate and dynamic real time hand gesture interactions and holographic object-manipulations [11]. Moreover, recent literature has reported the use of this technology for VR interactions [12] and in-car infotainment systems [13]. Finally, the underlying ultrasound modulation signals and waveforms have been investigated [14] such that rich haptic textures can be rendered in mid-air [15].

To the best of our knowledge, this paper presents the first integration of MR, wearable bio-sensing and mid-air haptics into an interactive prototype for multimodal (audio-visual and haptic) synchronized interaction with live (bio) holographic 3D objects. This demo enables users to see, touch and feel a hologram of a human heart beating as if it was their own, see Figure 1. The demo uses an Ultrahaptics display, a Magic Leap One MR headset, and an Apple Watch wristband, all synchronized, aligned and networked together such that updates from the wristband dynamically change the haptics and the animation speed of the beating heart. Interestingly, we observe that the inclusion of mid-air haptics has a strong impact on people's perception of holographic objects in comparison to visual stimuli alone. Specifically, early observations and user feedback revealed that participants manipulate and explore the haptic bio-holographic objects with much more care and agency.

Besides the novelty of our proposed haptic bio-hologram demo, the task of overcoming the integration complexities required to produce a functional multisensory experience in MR is the main contribution of this paper. In the following we detail a roadmap of how to implement such a system and our lessons learnt.

## 2 BACKGROUND

There is a long history of interactive holograms as well as a myriad of MR prototypes and demos. Many of the challenges still faced by the community were comprehensively captured by Bimber and Raskar [16]. Their review describes different display technologies and how information from environmental sensing can be used to enhance augmented synthetic reality. Going beyond single user interfaces, Billinghurst and Kato describe their vision of how MR could support computer supported collaborative work environments [17]. A number of studies explored hand gestures as a direct mean of interaction with holograms: HoloMuse [18] with applications in artefact manipulations in museum galleries; Yim et al. investigating interactions with geographical datasets using Microsoft HoloLens [19]; Grossman et al. exploring interactions with geometrical shapes using a spherical volumetric display [20]. Haptic holograms initially relied on force-feedback devices such as the Phantom stylus [21], vibrotactile gloves [22] or other wearables [8]. Breaking free from wearables and controllers, ultrasonic mid-air haptics was integrated with floating holographic displays in the HaptoClone and HaptoMime prototypes that used Asukanet glass plates [23, 24] and later with a Microsoft HoloLens in [11]. Both of these examples demonstrated the unique capability of mid-air haptics to instil a physical presence to otherwise intangible objects while freeing users' hands from any sort of instrumentation.

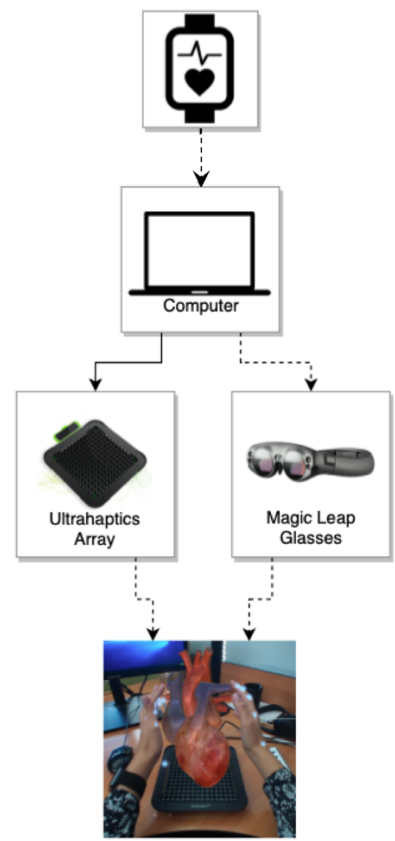

Figure 2: Schematic showing how the MR headset, the mid-air haptic device, and the heart rate monitor are connected together to produce touchless interactions with bio-holograms. Dashed arrows indicate a wireless connection.

## 3 DEMONSTRATION SETUP

Our demonstration setup is composed of a laptop PC wirelessly connected to a Magic Leap One MR headset and an Apple Watch. The PC is also wired to an Ultrahaptics STRATOS Explore (USX) device (see Figure 2). The USX is ergonomically located on a table or stand to let users easily access its interaction volume. Users can sit next the table setup and place their hands about 20 to 40cm above the USX to see, hear, and feel the holographic heart beating at the rate recorded by the Watch, see Figure 3. More advanced interactions such as 3D manipulation (rotation, translation, re-sizing) of the heart are possible but have been excluded from the demo to keep the duration of the overall experience under one minute, thereby maximizing the number of people who can experience the demo. A monitor displaying the MR headset viewpoint and a set of speakers can also be included to the setup. They would enable nearby people and passersby to also experience the audio-visual but not the haptic component of the prototype. The demo can be experienced unsupervised, but a presenter will be available to assist participants.

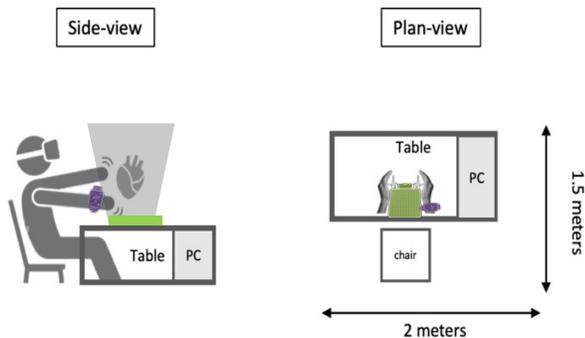

Figure 3: Schematic showing the demo setup and interaction space. The haptic interaction space is limited to that of the mid-air haptic device (approximately a volume of 40x40x60cm). The setup can be adjusted to allow for the user to stand and interact with the haptic holograms, while multiple mid-air haptic devices can be linked together in order to expand the interaction space.

## 4  KEY TECHNOLOGIES USED IN THE DEMO

The demo prototype uses a Magic Leap One MR headset for the holographic rendering of the heart, an Ultrahaptics STRATOS Explore Development kit for the delivery of mid-air haptics, and an Apple Watch wristband to measure the wearer's heartrate. The three components are synchronized and networked together by a laptop PC host. We now detail each of the components used in the demo highlighting any limitations, requirements, and options.

*MR Headset:* Magic Leap One is a standalone MR headset providing 40 degrees field of view and a resolution of 1280 x 960, which is sufficient to visualize medium sized objects in close proximity. The processing power of the device is close to standard gaming consoles with:

- CPU: NVIDIA® Parker SOC; 2 Denver 2.0 64-bit cores + 4 ARM Cortex A57 64-bit cores (2 A57's and 1 Denver accessible to applications)
- GPU: NVIDIA Pascal™, 256 CUDA cores; Graphic APIs: OpenGL 4.5, Vulkan, OpenGL ES 3.3+
- RAM: 8 GB.
- Storage Capacity: 128 GB

The wireless mobile form factor, in combination with high performance hardware and the custom Linux-based Lumin operating system, allows for wide applicability to spatial computing applications.

*Hand Tracking:* This demo does not use the hand tracking capabilities of the MR headset but instead, it uses a LEAP Motion controller that is attached to the Ultrahaptics display. This camera sensor is a small USB peripheral device that uses two monochromatic infrared cameras, three infrared LEDs, and machine vision proprietary software to track the user's hands and fingers at approximately 100 Hz refresh rate. Its field of view is 150 degrees wide and 120 degrees deep with an effective range of approximately 60 cm. The LEAP Motion is particularly good at tracking the user's palm position, orientation and motion, but is also capable to produce a full skeletal model of the hand and fingers. The latter is important as the skeletal joint 3D coordinates are used as inputs for the delivery of mid-air haptics.

*Mid-Air Haptic Feedback:* This demo uses the Ultrahaptics STRATOS Explore (USX) Development Kit. This phased array is composed of 16x16=256 ultrasonic transducers operating at 40 kHz with an effective ultrasonic focusing region of 40x40x60 cm$^3$ above the device, i.e. comparable to the LEAP Motion camera. Within this region, referred to as the *interaction volume*, the development kit can focus ultrasound at specific locations in 3D space and create a *focal point*. For a tactile sensation to be perceived however, an indentation of the skin due to the impinging acoustic radiation force is not enough. Further modulation of the carrier waves has to create shear waves in the skin tissue (i.e., a small localized skin vibrations) which are within the perceptible dynamic range of human skin ~5-500 Hz are needed. The focal point is be building block of two different types of haptic rendering methods: Amplitude Modulation (AM) [9], and spatiotemporal modulation (STM) [14]. Our prototype uses the STM method. The Ultrahaptics device comes with an API (written in C++ and C# with Unity3D Assets) which takes the focal point position coordinates and intensity as inputs. To place the focal point on a specific part of users' hands, e.g. at the centre of the palm or on the fingertips, the coordinates are obtained from the LEAP Motion API. Controlling the intensity allows to activate the haptic feedback when users' hands intersect with different parts of the holographic 3D object (in this case a heart organ). It also allows to temporally map the heartbeat to the haptic feedback. The complementarity of the LEAP Motion and Ultrahaptics API makes the *haptification* of mid-air interactions straight forward as long as they are *i)* detected by the LEAP Motion camera, and *ii)* within the focusing region of the USX.

*Heart Rate Sensor:* We read the user's heartrate from the PPG (photoplethysmogram) or ECG (electrocardiogram) sensor data stream obtained either via an Apple Watch (PPG) or a MAWI Band (ECG) device. The heart rate data from PPG sensor in Apple Watch is updated every 5 seconds and can be read from device by an iPhone paired to the Apple Watch via Bluetooth. The data is then transferred via WiFi-Direct to a local host computer for processing. After processing the data, our application updates both the visual appearance of the heart beating, and the haptic feedback characteristics bounded to the visuals. In the case of the ECG device, the workflow is similar to the Apple Watch and uses an Android phone that is paired to the MAWI band, reads data from it every 5 seconds via Bluetooth and transfers it via WiFi-Direct to the computer. While the ECG signal is richer in terms of heart beat features that could be visualized, the user needs to use both hands to take the heart rate measurements and therefore cannot interact with the mid-air bio-hologram at the same time. An alternative setup involves two users: one wearing the heartrate sensor, while the other wears the headset and interacts with the bio-hologram.

## 5 INTEGRATION AND IMPLEMENTATION

The goal of this demo is to visualize a virtual object in MR and allow the observer to "touch it" and feel haptic feedback. To this end, we have fixed in space the virtual object hovering over the Ultrahaptics display and have synchronized its appearance and the mid-air haptic feedback. Moreover, we investigated the temporal and spatial effects by allowing users to explore the holographic object with their hands and feel the heartbeat. The application was developed with the Unity game engine using additional Assets: OpenCV, Ultrahaptics binding for Unity3D, Lumin SDK, and a server application for the heart rate data stream transfer.

An important aspect of the application is the synchronization of Magic Leap One headset with the Ultrahaptics array. To this end, we have implemented a socket-to-socket interaction that synchronizes devices position on each rendering frame at 60 Hz.

Another important aspect of the demo protocol is the spatial scene calibration of the Magic Leap One camera module relative to the Ultrahaptics display position on the desk. The calibration can be achieved using either an OpenCV chessboard camera calibration approach or using the Magic Leap controller to place the virtual object graphics over the Ultrahaptics display.

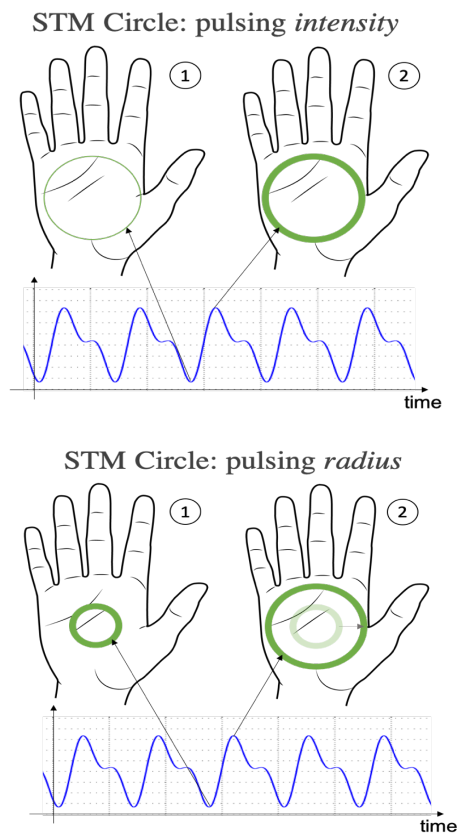

Figure 4: Schematic showing two haptic modalities (green) and a simplified data stream of a PPG signal (blue). Both haptic modalities are based on *STM Circle* haptics. The intensity of the PPG signal can be mapped to the intensity of the circle (top) or to its radius (bottom). For all the circles in the illustration, line thickness indicates haptic intensity.

Updates from the heart rate monitoring Apple Watch or MAWI Band are inputs of the haptic feedback modulation frequency and the animation speed of the beating heart. The heartrate signal is smoothed with a window averaging function and buffered for 6 seconds so that the haptic and visual playback feedback playback is representative of the heartbeat rate of the most recent reading. We have implemented two different haptic animations to map the heartrate data to the haptic sensation. The haptics are projected onto users' hands' only when they are intersecting with the holographic heart. The basis for the two haptic animations is a *STM Circle*. It consists of a single focal point (per interacting hand) that traverses a 3 cm radius circle at a rate of 100 Hz. Building on this circle sensation, we propose two approaches to produce a dynamic heartbeat sensation. The first one, called *pulsing intensity,* varies the haptic intensity according to the normalized power of the heart beat data stream. The second animation, called *pulsing radius,* changes the circle radius such that it increases and decreases in sync with the heartbeat. A schematic of the two animations is given in Figure 4. If the heart stops beating (i.e., flatlines), then both graphic and haptic animations of the holographic hearth become static. After internally trialling the two implementations, we concluded that the *pulsing radius* method was subjectively closer to what one would expect to feel when looking at the visual graphic animation of the beating holographic heart.

## 6 DEMO INTERACTION

The user interaction logically complements the Demo Setup shown in Figure 3. There are multiple ways users can interact with our demo. First, users observe in the headset the beating heart located just above the Ultrahaptics display. Synchronization of the headset and the Ultrahaptics position keeps the heart at the same position in physical space independently of viewing angles or distances. Second, users are able to influence the rate of the beating holographic heart. For example, by performing a few simple physical exercises, they will increase their heart rate. Similarly, the wearer of the sensor could perform some meditation or breathing exercises to calm themselves down thus reducing their heart rate. The increase/decrease of the heart rate will be reflected in both visual and haptic representations of the beating bio hologram. Third, when the user moves her hands over the holographic heart, she will observe a few markers that are super-imposed onto her hand joints as to visually indicate to the user that their hands are being tracked by the system – this indicates to the user that the system has recognised and is tracking her correctly and accurately. Fourth, if the user moves her hands towards the edge or inside the hologram to trigger the mid-air haptic feedback, she will feel the heart beat projected by the Ultrahaptics display.

## 7 DISCUSION AND CONCLUSION

In this paper we have explored the use of ultrasonic mid-air haptics for the untethered interaction with holographic 3D objects in MR environments with the aim of enhancing their physical perception. While most demos of mid-air haptics up until now have been limited to user interactions shown on a display screen or are purely immersed in a VR environment, here, users can directly interact with augmented holographic objects in a real-world environment using their bare hands. Going beyond the addition of mid-air haptics to 3D holograms, we have also experimented with the possibility of linking and influencing this multi-modal experience with a live data stream that corresponds to something as intimate and personal as someone's own heartbeat, motivated by haptic communication studies where empathy and emotions could be conveyed and enhanced via haptic devices over long distances [25][26]. Indeed, we found that this personalization has positively influenced user interaction and improved the overall experience.

Since the holograms supported by our prototype demonstrator can be seen, "touched" and felt, and also behave (graphically and haptically) according to a bio-sensed data, we have referred to them as *haptic bio-holograms*. To exemplify the different applications supported by our three-piece prototype system (MR headset, mid-air haptic device, and wearable bio-sensor) we have demonstrated a haptic bio-hologram of a human heart that beats at the rate provided by the bio-sensor. We argue that such multisensory experiences, coupled with next generation wireless connectivity and sensing hold many promising opportunities for future applications in medicine, entertainment, and education [27].

This prototype maps one input of interest, here the heartrate, to mid-air haptic feedback utilising a pulsing circle tactile pattern. There are many other possible ways to achieve this mapping. Future work could study the synthesis of different modalities and optimise for different use cases, whilst a deep perceptual study could help discern the perceptual parameters involved in MR mid-air haptic interactions.

## ACKNOWLEDGMENTS

Authors from Ultrahaptics acknowledge funding support from the EU's H2020 research and innovation programme under grant agreement No 801413, H-Reality.